\documentclass[aps,prc,twocolumn]{revtex4}
\usepackage{pdfpages}
\usepackage{graphicx}
\usepackage{amsmath}
\usepackage{booktabs}
\usepackage{multirow}
\usepackage{color}
\begin{document}

\title{Quark exchange effects in single flavored dibaryons}

\author{Cheng-Rong Deng$^{a}{\footnote{crdeng@swu.edu.cn}}$}
\affiliation{$^a$School of Physical Science and Technology,
Southwest University, Chongqing 400715, China}

\begin{abstract}
We reveal the quark exchange effects related to both the kinetic energy
and various interactions in the single flavored dibaryon bound states with
$^1S_0$ in the quark models. The hadron covalent bond can be established by
the shared identical quarks due to the quark exchange effect between two
colorless baryons. Such hadron covalent bond plays a decisive role in the
deuteronlike di-$\Omega_{ccc}$ and di-$\Omega_{bbb}$ covalent molecule states.
The $\sigma$-meson exchange is indispensable in the deuteronlike di-$\Delta^{++}$
and compact di-$\Omega$ states. The hadron covalent bond clearly appears in
the di-$\Delta^{++}$ state but is hidden in the di-$\Omega$ state. The
chromomagnetic interaction is always repulsive in the di-$\Delta^{++}$,
di-$\Omega$, di-$\Omega_{ccc}$, and di-$\Omega_{bbb}$ states. The
color-electric interaction is strongly attractive in the di-$\Omega$ state
but weakly attractive or repulsive in the di-$\Delta^{++}$, di-$\Omega_{ccc}$,
and di-$\Omega_{bbb}$ states.

\end{abstract}
\maketitle

\section{introduction}

The nuclear force is a residual color force among colorless nucleons, much
like the van der Waals forces among electric neutral molecules with the
exception of their energy scale. Its typical characters are the short-range
repulsion and medium-range attraction. It is a fundamental and
central subject of nuclear physics and has been intensively studied
since Yukawa proposed one pion exchange theory~\cite{Epelbaum:2008ga}.
With the developments of both experiment and computational physics,
one can generalize the nuclear force from the nucleon-nucleon ($NN$)
to other di-baryon systems involving strange, charm and bottom
flavors~\cite{Lee:2011rka,Gal:2016boi,Clement:2016vnl,Meng:2017fwb,
Gongyo:2017fjb,HALQCD:2018qyu,Junnarkar:2019equ,Richard:2020zxb,
Green:2021qol,Chen:2021hxs,Lyu:2021qsh,Mathur:2022ovu,Junnarkar:2022yak}.
The generalization is significantly important for describing the
nuclear force, nuclear structure and dense matter relevant to nuclear
physics and astrophysics~\cite{Shen:2019dls,Tong:2019juo,Drischler:2019xuo}.


The Fermi-Dirac statistics requires that identical fermions must be antisymmetrized
to satisfy the Pauli exclusion principle. In nuclear physics, the identical quark
exchange effect between different nucleons plays a critical role in the behaviors
of nuclei~\cite{Mulders:1987mw}. For example, most of European Muon Collaboration
effect can be attributed to the quark exchange effect between nucleons in
three-nucleon systems~\cite{Hoodbhoy:1986fn}. In molecular physics, the electrons
are shared by nuclei and their delocalization is an important effect contributing
to the formation of molecule covalent bond. Similarly, is there the hadron covalent
bond due to the shared identical quarks originating from the quark exchange
effects? The covalent hadron molecules were proposed, where the light identical
quarks are assumed to be shared by the heavy quarks~\cite{Du:2012wp,Chen:2021xlu}.
The hydrogen molecule-like $T^-_{bb}$ properly manifests such hadron covalent
bond~\cite{Deng:2021gnb,Maiani:2019lpu}.

In principle, the heavy identical quarks, if any, could also present such quark
exchange effects in the heavy hadron molecules as the light identical quarks do.
Admittedly, the heavy quark exchange effects are weaker than that of light identical
quarks because the exchange effects should be depressed by the large mass of the heavy
quarks. The single flavored dibaryons, di-$\Delta^{++}$, di-$\Omega$, di-$\Omega_{ccc}$,
and di-$\Omega_{bbb}$, cover from about 2.4 GeV to 30 GeV. Such a wide energy
region allows us to comprehensively address various dynamic mechanisms of the
low-energy strong interactions and their quark exchange effects. Technically,
the single flavored dibaryons possess the same flavor symmetry so that it is
convenient to perform a systematical investigation. In the channel with $^1S_0$,
the maximum attraction of the dibaryons is expected in comparison to other
channels because the Pauli exclusion principle between identical quarks at
short distances does not operate in this channel.

In this work, we attempt to systematically inspect the most promising single
flavored dibaryon bound states and figure out their binding energy and spatial
configuration from the perspective of quark models. More importantly, we prepare
to unveil such quark exchange effect and analyze various underlying binding
mechanisms in the dibaryon bound states very carefully.

After the introduction, the paper is organized as follows. In Sec. II we describe
the quark models for nuclear force. In Sec. III we briefly introduce the trial wave
functions for ground state baryons and dibaryons. In Sec. IV we present the
numerical results and discussions. In the last section we list a brief summary.

\section{quark models for nuclear force}

The strong interactions are widely described by quantum chromodynamics (QCD)
in the standard model of particle physics. However, the {\it ab initio}
calculation of the hadron spectroscopy and the hadron-hadron interaction
directly from QCD is very difficult due to the complicated nonperturbative
natures. Therefore, the QCD-inspired constituent quark model is a powerful
implement in obtaining physical insight for these complicated strong
interacting systems. We apply naive quark model (NQM) and chiral quark model
(ChQM) of the Salamanca group to investigate the single flavored dibaryons in
this work. Those models were developed based on the reasonable description
of the natures of baryons and the $NN$ interactions.

\subsection{Naive quark model}

Naive quark model generally includes an effective one-gluon-exchange
(OGE) potential $V^{\rm oge}$ directly coming from the OGE diagram in
QCD~\cite{DeRujula:1975qlm} and an artificial quark confinement potential
$V^{\rm con}$. The model can provide a very good description of the light
baryons~\cite{Isgur:1978xj,Isgur:1979be}. In the $NN$ interactions, the
model can obtain the short-range repulsive core by the spin-spin part
of the inter quark interaction between nucleons and the Pauli exclusion
principle enforced by the quark structure of the nucleon~\cite{Liberman:1977qs}.
However, the medium-range attraction is absent~\cite{Neudachin:1977vt,Fujiwara:1985ze}.
The model hamiltonian used in this work reads
\begin{eqnarray}
\begin{aligned}
&H_n=\sum_{i=1}^n \left(m_i+\frac{\mathbf{p}_i^2}{2m_i}
\right)-T_{\rm cm}+\sum_{i<j}^n (V_{ij}^{\rm oge}+V_{ij}^{\rm con}), \\
\noalign{\smallskip}
&V_{ij}^{\rm oge}={\frac{\alpha_{s}}{4}}\boldsymbol{\lambda}_{i}
\cdot\boldsymbol{\lambda}_{j}\left({\frac{1}{r_{ij}}}-
{\frac{\boldsymbol{\sigma}_{i}\cdot\boldsymbol{\sigma}_{j}}
{6m_im_jr_0^2(\mu_{ij})r_{ij}}}e^{-\frac{r_{ij}}{r_0(\mu_{ij})}}\right),\\
\noalign{\smallskip}
&V^{\rm con}_{ij}=-a_c\boldsymbol{\lambda}_i\cdot\boldsymbol{\lambda}_jr^2_{ij}.\\
\noalign{\smallskip}
\end{aligned}
\end{eqnarray}
$m_i$ and $\mathbf{p}_i$ are the mass and momentum of the quark $q_i$, respectively.
$T_{\rm cm}$ is the center-of-mass kinetic energy. $\boldsymbol{\lambda}_{i}$ and
$\boldsymbol{\sigma}_{i}$ stand for the SU(3) Gell-Mann matrices and SU(2) Pauli
matrices, respectively. $r_{ij}$ is the distance between two quarks $q_i$ and $q_j$
and $\mu_{ij}$ is their reduced mass, $r_0(\mu_{ij})=\frac{\hat{r}_0}{\mu_{ij}}$.
The quark-gluon coupling constant $\alpha_s$ adopts an effective scale-dependent
form,
\begin{equation}
\alpha_s(\mu_{ij})=\frac{\alpha_0}{\ln\frac{\mu_{ij}^2}{\Lambda_0^2}}.
\end{equation}
The model parameters $a_c$, $\hat{r}_0$, $\Lambda_0$, and $\alpha_0$ can be determined
by fitting the ground state baryon spectrum.

\subsection{Chiral quark model}

To achieve medium- and long-range behaviors of nuclear force, the hybrid quark
model was established by introducing one $\pi$-meson exchange and one $\sigma$-meson
exchange on the baryon level~\cite{Oka:1982qa}. The effective meson-exchange
potential between two nucleons were considered to simulate the effects of the
meson cloud surrounding the quark core. In this way, the implementation of
chiral symmetry at the quark potential level was needed for the sake of
consistency~\cite{Manohar:1983md}. The constituent quark mass appears because
of the spontaneous breaking of the chiral symmetry at some momentum scale.
Once a constituent quark mass is generated, such quarks have to interact
through Goldstone bosons. $\sigma$-meson as well as $\pi$-meson exchanges
on the quark level were introduced in the NQM, i.e. SU(2) ChQM. The model
can well describe the hadron spectra, $NN$ phase shifts and the deuteron
~\cite{Obukhovsky:1990tx,Fernandez:1993hx,Valcarce:1995dm}. Subsequently,
the extended model, SU(3) ChQM, was employed to investigate the nucleon-hyperon
and hyperon-hyperon interactions~\cite{Fujiwara:1996qj}. In the light quark
sector ($u$, $d$ and $s$), the meson-exchange interactions $V_{ij}^{\pi}$,
$V_{ij}^{K}$, $V_{ij}^{\eta}$, and $V_{ij}^{\sigma}$ are included and the
relative parameters are taken from Ref.~\cite{Vijande:2004he}. Note that the
vector meson exchange interactions are excluded to avoid the possible double
counting of the short-range repulsion in the model study of the baryon-baryon
interactions~\cite{Valcarce:2005em}. In the heavy quark sector ($b$ and $c$),
the meson-exchange interaction does not happen because the chiral symmetry
is explicitly broken.

\section{wave functions}

The wave function of ground state baryons with isospin $I$ and angular momentum
$J$ can be written as the direct products of color part $\chi_c$, isospin-spin
part $\eta_{is}$ and spatial part $\psi$,
\begin{eqnarray}
\Phi^{B}_{IJ}=\chi_c\otimes\eta_{is}\otimes\psi.
\end{eqnarray}
The spin-flavor symmetry ${\rm SU_{sf}(6)\supset SU_{s}(2)\otimes SU_{f}(3)}$
is taken into account in the SU(3) ChQM. The spin-flavor symmetry
${\rm SU_{sf}(4)\supset SU_{s}(2)\otimes SU_{f}(2)}$ is involved in the NQM.
The color singlet $\chi_c$ is antisymmetrical so that the spatial $\psi$ must be
symmetrical for identical quarks in the ground state baryons.

We define a set of Jacobi coordinates $\mathbf{r}_{ij}$, $\mathbf{r}_{ijk}$ and
$\mathbf{R}_{\rm cm}$,
\begin{eqnarray}
\begin{aligned}
&\mathbf{r}_{ij}=\mathbf{r}_i-\mathbf{r}_j,\ {} \ {}
\mathbf{r}_{ijk}=\frac{m_i\mathbf{r}_i+m_j\mathbf{r}_j}{m_i+m_j}-\mathbf{r}_k,
\end{aligned}
\end{eqnarray}
$\mathbf{R}_{\rm cm}$ stands for the center of mass of baryons.
In the center of mass frame, the symmetrical spatial wave functions of
baryons composed of three identical particles can be expressed as
\begin{eqnarray}
\psi=\phi(\mathbf{r}_{12})\phi(\mathbf{r}_{123})+
\phi(\mathbf{r}_{13})\phi(\mathbf{r}_{132})+\phi(\mathbf{r}_{23})\phi(\mathbf{r}_{231}).
\end{eqnarray}
For baryons with only two identical particles, we just consider their antisymmetry
in the simplest way because we pay more attention to the residual interaction between
two colorless baryons than the properties of the individual baryon. The spatial
wave function can be written as
\begin{eqnarray}
\psi=\phi(\mathbf{r}_{12})\phi(\mathbf{r}_{123}),
\end{eqnarray}
where the quarks $1$ and $2$ are identical particles. For baryons with three different
particles, the spatial wave function is also taken as Eq. (6), where the quarks $1$
and $2$ are the two light quarks. In fact, the influence of the simplification on
the baryon is not obvious in comparison of the case including the Jacobi coordinate
in Eq. (5).

Accurate model calculations are a primary requirement for the exact understanding
the properties of dibaryons. The Gaussian expansion method (GEM) has been proven
to be rather powerful to solve the few-body problem in nuclear physics~\cite{Hiyama:2003cu}.
According to the GEM, the relative motion wave functions $\phi(\mathbf{x})$ is
expanded as the superpositions of a set of Gaussian functions with different
sizes,
\begin{eqnarray}
\phi(\mathbf{x})&=&\sum_{n=1}^{n_{max}}c_nN_{nl}x^{l}
e^{-\nu_nx^2}Y_{lm}(\hat{\mathbf{x}}),
\end{eqnarray}
where $\mathbf{x}$ represents $\mathbf{r}_{ij}$ and $\mathbf{r}_{ijk}$. More details
about the GEM can be found in Ref.~\cite{Hiyama:2003cu}.

The wave function of the ground state dibaryons with defined isospin-spin can
be expressed as
\begin{eqnarray}
\Psi_{IJ}^{\rm Dibaryon}=\sum_{\xi}
c_{\xi}\mathcal{A}\left\{\left[\Phi^{\rm B_1}_{I_1J_1C_1}
\Phi^{\rm B_2}_{I_2J_2C_2}\right]_{IJ}F(\boldsymbol{\rho})\right\},
\end{eqnarray}
where $\Phi^{\rm B_1}_{I_1J_1C_1}$ and $\Phi^{\rm B_2}_{I_2J_2C_2}$ are the wave functions
of the individual baryon and the subscripts $C_i$ denote their color representations.
In principe, the dibaryons should be the mixture of color singlet and hidden color octet.
Here, we mainly focus on the quark exchange effect between two colorless baryons similar
to the chemical covalent bond. The hidden color effect is left for the future work.
$\mathcal{A}$ is antisymmetrization operator acting on the identical quarks
belonging to two different baryons. $\xi$ stands for all possible isospin-spin-color
combinations $\{I_1,I_2,J_1,J_2,C_1,C_2\}$ that can be coupled into the quantum numbers
of the dibaryon. The coefficients $c_{\xi}$ can be determined by the dynamics of the
dibaryon. $F(\boldsymbol{\rho})$ is the relative motion wave function between two
baryons and can also be expanded by a set of Gaussian bases.

\section{numerical results and discussions}
\subsection{Model parameters and baryon spectra}

The $u$- and $d$-quark mass $m_{u,d}$ is taken to be one third of that of nucleon.
With the Minuit program~\cite{James:1975dr}, other model parameters can be determined
by fitting ground state baryon spectrum by accurately solving the three-body Schr\"{o}dinger
equation. The parameters and ground state baryon spectrum are presented in
Tables~\ref{parameters} and~\ref{baryons}, respectively.

\begin{table}[ht]
\caption{Model parameters. Quark masses and $\Lambda_0$ unit in MeV, $a_c$ unit in
MeV$\cdot$fm$^{-2}$, $r_0$ unit in MeV$\cdot$fm and $\alpha_0$ is dimensionless.}\label{parameters}
\begin{tabular}{lccccccccccccccccc}
\toprule[0.8pt] \noalign{\smallskip}
Parameter&$m_{u,d}$~~&~$m_{s}$~&~~~$m_c$~~~&~~$m_b$~~&~~~$a_c$~~~&~$\alpha_0$~&~~~$\Lambda_0$~~~&~$r_0$~   \\
\noalign{\smallskip}
\toprule[0.8pt] \noalign{\smallskip}
NQM     & 313~ & 450 & 1633  & 4991 & 118  & 3.03 & 67.7 & 90.8  \\
\noalign{\smallskip}
ChQM    & 313~ & 500 & 1614  & 4982 & 45.6 & 3.76 & 21.9 & 95.7  \\
\noalign{\smallskip}
\toprule[0.8pt]
\end{tabular}
\end{table}

In addition, we calculate the mass root-mean-square (rms) radius of quark core of baryons
with their eigenvectors. The mass rms radius was defined as~\cite{Silvestre-Brac:1985aip,
Silvestre-Brac:1996myf}
\begin{eqnarray}
\langle\mathbf{r}^2\rangle^{\frac{1}{2}}=\left (\sum_{i=1}^3\frac{m_i\langle(\mathbf{r}_i-
\mathbf{R}_{\rm cm})^2\rangle}{m_1+m_2+m_3}\right )^{\frac{1}{2}}.
\end{eqnarray}
We list the numerical results in Table~\ref{baryons}, which are close to those in
Refs~\cite{Silvestre-Brac:1985aip,Silvestre-Brac:1996myf}. The mass rms radius is not an
observable, but it is nevertheless a very interesting quantity, which gives the size of the
baryons in the constituent quark models. In general, the mass rms radius of quark core is
smaller than physical radius of baryons because the contributions from the meson cloud
surrounding the valence quarks are not included in the model calculations.

\begin{table*}[ht]
\caption{Mass spectra of baryon ground states unit in MeV and mass rms radius of quark
core unit in fm. PDG is the abbreviation of particle data group. The ``$\times$" denotes
that the state does not exist in the experiment.}\label{baryons}
\begin{tabular}{lcccccccccc} \toprule[0.8pt] \noalign{\smallskip}
\noalign{\smallskip}
\multirow{2}{*}{Baryon}&\multirow{2}{*}{~~$I(J^P)$~~}&NQM&ChQM&PDG&\multirow{2}{*}{Baryon}
&\multirow{2}{*}{~~$I(J^P)$~~}&NQM&ChQM&PDG&\\
&&~~Mass,~Radius~~&~~Mass,~Radius~~&~~Mass~~&&&~~Mass,~Radius~~&~~~Mass,~Radius~~~&Mass   \\
\noalign{\smallskip}
\toprule[0.8pt]
$\Delta(1232)$        &$\frac{3}{2}(\frac{3}{2}^+)$&1234,~0.51&1242,~0.64&1232&$\Omega_c(2770)^{0}$  &$0(\frac{3}{2}^+)$          &2768,~0.38 &2751,~0.48 &2766 \\
\noalign{\smallskip}
$\Sigma^*(1385)$      &$1(\frac{3}{2}^+)$          &1393,~0.50&1391,~0.61&1385&$\Xi^{++}_{cc}(3622)$ &$\frac{1}{2}(\frac{1}{2}^+)$&3635,~0.33 &3636,~0.43 &3622 \\
\noalign{\smallskip}
$\Xi^*(1530)$         &$\frac{1}{2}(\frac{3}{2}^+)$&1537,~0.48&1521,~0.58&1530&$\Lambda_b^0(5620)$   &$0(\frac{1}{2}^+)$          &5624,~0.23 &5607,~0.29 &5620 \\
\noalign{\smallskip}
$\Omega(1672)$        &$0(\frac{1}{2}^+)$          &1668,~0.47&1653,~0.55&1672&$\Sigma_b(5810)$      &$1(\frac{1}{2}^+)$          &5810,~0.24 &5814,~0.31 &5808 \\
\noalign{\smallskip}
$N(939)$              &$\frac{1}{2}(\frac{1}{2}^+)$&942,~0.47 &938,~0.54 &939 &$\Sigma_b^*(5830)$    &$1(\frac{3}{2}^+)$          &5838,~0.24 &5826,~0.32 &5830 \\
\noalign{\smallskip}
$\Sigma(1192)$        &$1(\frac{1}{2}^+)$          &1178,~0.46&1206,~0.54&1192&$\Xi_b(5792)$         &$\frac{1}{2}(\frac{1}{2}^+)$&5790,~0.24 &5816,~0.30 &5790 \\
\noalign{\smallskip}
$\Xi(1315)$           &$\frac{1}{2}(\frac{1}{2}^+)$&1321,~0.45&1336,~0.50&1315&$\Xi^{\prime}_b(5935)$&$\frac{1}{2}(\frac{1}{2}^+)$&5927,~0.25 &5937,~0.32 &5935 \\
\noalign{\smallskip}
$\Lambda(1116)$       &$0(\frac{1}{2}^+)$          &1121,~0.46&1109,~0.51&1116&$\Xi_b(5955)$         &$\frac{1}{2}(\frac{3}{2}^+)$&5955,~0.25 &5949,~0.32 &5955 \\
\noalign{\smallskip}
$\Lambda_c^+(2286)$   &$0(\frac{1}{2}^+)$          &2288,~0.35&2270,~0.44&2285&$\Omega_b^-(6046)$    &$0(\frac{1}{2}^+)$          &6052,~0.25 &6064,~0.32 &6046 \\
\noalign{\smallskip}
$\Sigma_c(2455)$      &$1(\frac{1}{2}^+)$          &2440,~0.36&2463,~0.47&2455&$\Xi^{++}_{cc}$       &$\frac{1}{2}(\frac{3}{2}^+)$&3718,~0.34 &3667,~0.44 &$\times$ \\
\noalign{\smallskip}
$\Sigma_c(2520)$      &$1(\frac{3}{2}^+)$          &2517,~0.37&2493,~0.48&2520&$\Xi^{0}_{bb}$        &$\frac{1}{2}(\frac{1}{2}^+)$&10244,~0.23&10264,~0.29&$\times$ \\
\noalign{\smallskip}
$\Xi_c(2467)$         &$\frac{1}{2}(\frac{1}{2}^+)$&2462,~0.36&2485,~0.44&2466&$\Xi^{0}_{bb}$        &$\frac{1}{2}(\frac{3}{2}^+)$&10277,~0.23&10277,~0.29&$\times$ \\
\noalign{\smallskip}
$\Xi^{\prime}_c(2578)$&$\frac{1}{2}(\frac{1}{2}^+)$&2566,~0.36&2591,~0.46&2578&$\Omega_{ccc}$        &$0(\frac{3}{2}^+)$          &4881,~0.32 &4791,~0.39 &$\times$ \\
\noalign{\smallskip}
$\Xi_c(2645)$         &$\frac{1}{2}(\frac{3}{2}^+)$&2641,~0.37&2622,~0.48&2645&$\Omega_{bbb}$        &$0(\frac{3}{2}^+)$          &14666,~0.21&14662,~0.25&$\times$ \\
\noalign{\smallskip}
$\Omega_c(2695)^0$    &$0(\frac{1}{2}^+)$          &2698,~0.36&2721,~0.46&2695\\

\noalign{\smallskip}
\toprule[0.8pt]
\end{tabular}
\end{table*}

\subsection{Natures of di-$\Delta^{++}$, di-$\Omega$, di-$\Omega_{ccc}$, and di-$\Omega_{bbb}$}

{\it Binding energies}. Using the well defined trial wave function, we can obtain the
eigenvalue and eigenvector of the single flavored dibaryons with $^1S_0$ by accurately
solving the six-body Schr\"{o}dinger equation in the quark models. Subsequently, we can
arrive at their binding energy $E_b=E_6(\rho)-E_6(\infty)$, where $E_6(\rho)$ denotes
the minimum of the dibaryons at the average separation $\rho$ between two baryons and
$E_6(\infty)$ is the mass of two isolated baryons in the models. Such a subtraction
procedure can greatly reduce the influence of the inaccurate model parameters and
hadron spectra on the binding energy, which is properly exhibited in study of the
deuteronlike molecular state $T^+_{cc}$~\cite{Deng:2021gnb}. To illustrate the formation
mechanism of the bound dibaryons, we calculate and decompose the contribution to $E_b$
from each part of the model Hamiltonian. We present the binding energy and various
contributions in Table~\ref{dibaryons}.
\begin{table*}[ht]
\caption{Binding energy $E_b$ and the contribution of each part in the Hamiltonian to $E_b$,
$\Delta V^{\rm con}$, $\Delta V^{\rm coul}$, $\Delta V^{\rm cm}$, $\Delta T$, $\Delta V^{\sigma}$,
$\Delta V^{\pi}$, and $\Delta V^{\eta}$ are confinement term, Coulomb term, chromomagnetic term,
kinetic energy, $\sigma$-, $\pi$-, and $\eta$-meson exchange term, respectively, unit
in MeV. $\langle\mathbf{r}^2\rangle^{\frac{1}{2}}$ is the size of a single baryon,
$\langle\boldsymbol{\rho}^2\rangle^{\frac{1}{2}}$ is the distance between two baryons
and $d$ is the distance predicted by the Heisenberg uncertainty-relation formula,
unit in fm.} \label{dibaryons}
\begin{tabular}{ccccccccccccccccccccccc}
\toprule[0.8pt] \noalign{\smallskip}
Dibaryon&~Model~&~~~~~~~$E_b$~~~~~~~&$\Delta V^{\rm con}$&~~~~~$\Delta V^{\rm coul}$~~~~~&$\Delta V^{\rm cm}$&~~~~~$\Delta T$~~~~~
&$\Delta V^{\sigma}$&~~~~$\Delta V^{\pi}$~~~~&~$\Delta V^{\eta}$~&~~$\langle\mathbf{r}^2\rangle^{\frac{1}{2}}$~~
&~$\langle\boldsymbol{\rho}^2\rangle^{\frac{1}{2}}$~&&$d$\\
\noalign{\smallskip}\toprule[0.8pt] \noalign{\smallskip}
\multirow{2}{*}{di-$\Delta^{++}$}  & NQM  & Unbound  &          &          &       &         &           &       &       & 0.51 & $\infty$ && $\infty$ \\
\noalign{\smallskip}
                                   & ChQM & $-7.57$  & $-2.99$  &  $0.69$  & 12.79 & $-4.14$ & $-27.42$  & 12.12 & 1.36  & 0.64 &   2.48   &&  ~2.04~  \\
\noalign{\smallskip}
\noalign{\smallskip}
\multirow{2}{*}{di-$\Omega$~~~~}   & NQM  & Unbound  &          &          &       &         &           &       &       & 0.47 & $\infty$ \\
\noalign{\smallskip}
                                   & ChQM & $-61.66$ & $-18.93$ & $-20.72$ & 33.99 & $44.35$ & $-116.97$ & 0.00  & 16.62 & 0.55 &   1.03   &&  0.61  \\
\noalign{\smallskip}
\noalign{\smallskip}
\multirow{2}{*}{di-$\Omega_{ccc}$} & NQM  & $-0.54$  &  $0.13$  &  $1.91$  & 2.56  & $-5.14$ &           &       &       & 0.32 &   3.71   &&  3.84  \\
\noalign{\smallskip}
                                   & ChQM & $-1.16$  & $-0.28$  & $1.08$   & 2.44  & $-4.40$ &           &       &       & 0.39 &   2.34   &&  2.65  \\
\noalign{\smallskip}
\noalign{\smallskip}
\multirow{2}{*}{di-$\Omega_{bbb}$} & NQM  & $-1.07$  & $-0.25$  & $-0.25$  & 1.30  & $-1.87$ &           &       &       & 0.21 &   1.96   &&  1.57  \\
\noalign{\smallskip}
                                   & ChQM & $-1.08$  & $0.05$   & $-0.07$  & 1.18  & $-2.24$ &           &       &       & 0.25 &   1.80   &&  1.57  \\
\noalign{\smallskip}
\toprule[0.8pt]
\end{tabular}
\end{table*}

Oka {\it et al} found that the di-$\Delta^{++}$ state with $^1S_0$ can not be bound in the
similar NQM~\cite{Oka:1980ax}, which is strengthened by the present work. The di-$\Delta^{++}$
state can establish a shallow bound dibaryons with a binding energy about 8 MeV in the ChQM.
The previous ChQM studies on the state indicated that it is a deep bound state with a binding
energy about 10 to 50 MeV~\cite{Li:2000cb,Huang:2013nba}. Quark delocalization and color
screening model, where the $\sigma$-meson exchange effect is replaced with a hybrid confinement
potential and quark delocalization, also gave similar results~\cite{Huang:2013nba,Pang:2001xx}.
In one word, all of the models that provide the intermediate range attraction of nuclear force
support the existence of the bound di-$\Delta^{++}$ state. Exactly, the di-$\Delta^{++}$ state
is a resonance rather than a bound state in the quark models because it can decay into the
$pp\pi^+\pi^+$ channel.

In the NQM, the di-$\Omega$ state with $^1S_0$ is unbound because of the absence of the
binding mechanism. However, it becomes a deep bound state with a binding energy of about 62
MeV in the ChQM owing to the strongly $\sigma$-meson exchange. Other versions of SU(3) ChQM
also preferred the deep bound di-$\Omega$ state and its binding energy is around 80-120
MeV~\cite{Zhang:2000sv,Li:2000cb,Huang:2019hmq}. Recently, lattice QCD predicted that the
binding energy of the di-$\Omega$ state is about $1.6(6)(^{+0.7}_{-0.6})$ MeV with a large
volume and nearly physical pion mass~\cite{Gongyo:2017fjb}. Subsequently, the quark
delocalization and color screening model and QCD sum rule also suggested the existence
of a loosely molecular di-$\Omega$ state~\cite{Huang:2019hmq,Chen:2019vdh}. Comparatively
speaking, the ChQMs provide the strongly attraction for the di-$\Omega$ state due to the
$\sigma$-meson exchange, which may be pushed down by the introduction of the vector meson
exchanges. The vector meson exchanges were used to reduce the strongly attraction also
induced by the $\sigma$-exchange in the doubly heavy state $T^+_{cc}$~\cite{He:2023ucd}.

With regard to the fully heavy quark systems, the NQM and ChQM do not exist any dissimilarities
except for their model parameters in this work. The di-$\Omega_{ccc}$ and di-$\Omega_{bbb}$
states with $^1S_0$ can establish very shallow bound states with a binding energy around 1 MeV.
Quark delocalization and color screening model also gave similar results~\cite{Huang:2020bmb}.
Therefore, the shallow di-$\Omega_{ccc}$ and di-$\Omega_{bbb}$ bound states seem to be
independent of quark models. The extended one-boson-exchange model including heavy meson
exchange prefers to describe the di-$\Omega_{ccc}$ and di-$\Omega_{bbb}$ as shallow bound
states~\cite{Liu:2021pdu}. In the lattice QCD, the di-$\Omega_{ccc}$ is a loose bound
state~\cite{Lyu:2021qsh} while the di-$\Omega_{bbb}$ prefers a very deep bound
state~\cite{Mathur:2022ovu}.

{\it Spatial configurations}.
We can precisely calculate the average distance, $\langle\boldsymbol{\rho}^2\rangle^{\frac{1}{2}}$
in Table~\ref{dibaryons}, between two baryons with the eigenvector. Combining
the average distance with the mass rms radius of baryons, we figure out the spatial
configuration of the dibaryon bound states. In the di-$\Delta^{++}$, di-$\Omega_{ccc}$,
and di-$\Omega_{bbb}$ states, the average distances $\langle\boldsymbol{\rho}^2\rangle^{\frac{1}{2}}$
are obviously larger than the sum of the mass rms radius $\langle\mathbf{r}^2\rangle^{\frac{1}{2}}$
of the corresponding baryons. They are deuteronlike states because two baryons are very
far apart from each other and do not overlap entirely. The di-$\Omega$ state is a compact
state rather than a loose deuteronlike state because two $\Omega$s are partly overlapped
from its $\langle\mathbf{r}^2\rangle^{\frac{1}{2}}$ and $\langle\boldsymbol{\rho}^2\rangle^{\frac{1}{2}}$,
which is supported by Ref.~\cite{Zhang:2000sv}. If taking into account the contributions
from meson cloud surrounding the valence quarks to the size of $\Omega$, two $\Omega$s
are strongly overlapped in the di-$\Omega$ state.

In general, the average distance between two baryons is related to the binding energy
$E_b$ of the dibaryon states. One can therefore roughly estimate the distance between
two completely separated baryons by the Heisenberg uncertainty-relation formula~\cite{Bignamini:2009sk},
\begin{eqnarray}
d\approx\frac{\hbar c}{\sqrt{2\mu E_b}},
\end{eqnarray}
where $\mu$ is the reduced mass of two baryons. This formula was proposed to roughly
estimate the size of the state $X(3872)$ described as a $D^0\bar{D}^{*0}$
molecule~\cite{Bignamini:2009sk}. For the deuteron, one can verify that the formula
is effective. For the deuteronlike di-$\Delta^{++}$, di-$\Omega_{ccc}$ and di-$\Omega_{bbb}$
states, the differences between $\langle\boldsymbol{\rho}^2\rangle^{\frac{1}{2}}$
and $d$ are obviously smaller than the sizes of the deuteronlike states. For the compact
di-$\Omega$ state, the difference is 0.42 fm so that it cannot be ignored relative to
the size predicted by the formula. The di-$\Delta$ resonance $d^*(2380)$ reported by
the WASA-at-COSY Collaboration is very similar to the di-$\Omega$ state because
both of them are deeply bound states~\cite{Bashkanov:2008ih}. However, the reliable
information about the spatial configuration of the state $d^*(2380)$ is unavailable
so far~\cite{Dong:2023xdi}. The reliability of this formula is an open question in
the estimating the size of compact multiquark states.

\subsection{Quark exchange effects and binding mechanisms}

{\it Chromomagnetic and color-electric interactions}.
Both the chromomagnetic and color-electric interactions depend on the color
factor $\langle\boldsymbol{\lambda}_{i}\cdot\boldsymbol{\lambda}_{j}\rangle$
so that their contributions to the binding energy come from the quark exchange
effects between two colorless objects. From Table~\ref{dibaryons}, one can
see that the chromomagnetic interaction provide some repulsions in all of the
bound single flavored dibaryons predicted by our models. The repulsion is in
the order of tens MeV in the di-$\Delta^{++}$ and di-$\Omega$ states but less
than 3 MeV in the di-$\Omega_{ccc}$ and di-$\Omega_{bbb}$ states due to the
large mass of heavy quarks. The contributions from the color-electric interaction,
i.e. the color Coulomb plus color confinement, are small in the deuteronlike
di-$\Delta^{++}$, di-$\Omega_{ccc}$, and di-$\Omega_{bbb}$ states. The reason
is that the Coulomb interaction is inverse proportional to the distance and the
effective interacting range of the confinement potential is around 1 fm. For
the same reason, the color-electric interaction provides a stronger attraction
in the compact di-$\Omega$ state. On the whole, the chromomagnetic and color-electric
interactions can just provide a small quantity of attractions even a few of
repulsions. In this way, none of bound single flavored dibaryons can be produced
completely by means of the chromomagnetic and color-electric interactions, which
approves the conclusion about the stability of fully heavy dibaryons in the
extended chromomagnetic model~\cite{Weng:2022ohh}.

{\it Meson exchange interactions}.
The meson exchange interactions are independent of colors. Their contributions
to the binding energy come from both the direct term (main) and the quark exchange
effects. The $\sigma$-meson exchange provides a strongly attraction in the both
di-$\Delta^{++}$ and di-$\Omega$ states with $^1S_0$ in the ChQM while the $\pi$-
and $\eta$-meson exchanges are repulsive in the states. The total contribution
from the $\sigma$-, $\pi$-and $\eta$-meson exchange is attractive. Exactly
similar to the deuteron, the $\sigma$-meson exchange plays a predominant role
in the formation of the di-$\Delta^{++}$ and di-$\Omega$ states with $^1S_0$ in
the ChQM. The absence of $\sigma$-meson exchange in the NQM directly leads to the
disappearance of the di-$\Delta^{++}$ and di-$\Omega$ bound states. The one boson
exchange model based on the nuclear force was extended to predict the existence
of di-$\Omega_{ccc}$ and di-$\Omega_{bbb}$ by introducing charmonium and bottomonia
exchange potential~\cite{Liu:2021pdu}. In this work, the di-$\Omega_{ccc}$
and di-$\Omega_{bbb}$ states can establish bound states independence of any meson
exchanges. That is to say, the meson exchanges in heavy quark sector are not
indispensable in the formation of the dibaryon bound states, which implies that
there may exist some novel binding mechanism.

{\it Hadron covalent bond}.
Assuming the size of baryons does not change obviously in their interaction,
the kinetic energy contribution $\Delta T$ to the binding energy is the sum
of the relative motion part between two baryons and the exchange kinetic term
introduced by exchanging identical quarks. The study on the nucleon-nucleon
system indicated that the exchange kinetic term can reduce the total kinetic
energy, i.e. the term is negative~\cite{Nzar:1990ci}. Hoodbhoy and Jaffe
pointed out that the reduction is equivalent to a softening of the quark
momentum distribution~\cite{Hoodbhoy:1986fn}.

In the dibaryon systems, the identical quark exchange permits a quark in
one baryon to roam into the other baryon, which can effectively expand the
Hilbert space of the systems. The delocalized identical quarks are shared
by the dibaryon so that the hadron covalent bond similar to the molecular
one can establish. The most intuitive representation of such hadron covalent
bond is the reduction of the total kinetic energy of the system because of
the Heisenberg uncertainty relation. In other words, the hadron covalent bond
can provide an effective binding mechanism. As can be seen from $\Delta T$ in
Table III, the effect of the hadron covalent bond in the deuteronlike
di-$\Delta^{++}$, di-$\Omega_{ccc}$, and di-$\Omega_{bbb}$ states conspicuously
emerge because of the small relative motion energy between two remarkably
separated baryons. However, the effect in the compact di-$\Omega$ state is
hidden by the larger relative motion kinetic energy between two overlapped
$\Omega$s.

In the di-$\Delta^{++}$ and di-$\Omega$ states, the main binding mechanism
is the $\sigma$-meson exchange or its alternative effect while the hadron
covalent bond is secondary. In strong contrast, the absolute predominant
binding mechanism in the di-$\Omega_{ccc}$ and di-$\Omega_{bbb}$ states is
the hadron covalent bond so that we can call di-$\Omega_{ccc}$ and
di-$\Omega_{bbb}$ bound states the covalent hadron molecules. Note that
the large mass of heavy quarks depresses the repulsive chromomagnetic
interaction, which is beneficial to establish the covalent hadron molecules.
\begin{table}[ht]
\caption{Dependence of binding energy $E_b$ and various contributions to $E_b$
on the heavy quark mass, unit in MeV.}\label{dependence}
\begin{tabular}{lccccccccccccccccccccccc}
\toprule[0.8pt] \noalign{\smallskip}
~$M_Q$~&~~Model~~&~~~~$E_b$~~~~&~~$\Delta V^{\rm con}$~~&~~$\Delta V^{\rm coul}$~~&~~$\Delta V^{\rm cm}$~~&~~$\Delta T$   \\
\noalign{\smallskip}
\toprule[0.8pt] \noalign{\smallskip}
\multirow{2}{*}{1500}    & NQM  & $-0.46$ & $0.09$  & $1.81$  &  $2.55$   & $-4.90$ \\
\noalign{\smallskip}
                         & ChQM & $-1.11$ & $-0.31$ & $1.10$  &  $2.50$   & $-4.40$ \\
\noalign{\smallskip}
\multirow{2}{*}{2000}    & NQM  & $-0.65$ & $0.21$  & $1.97$  &  $2.42$   & $-5.26$  \\
\noalign{\smallskip}
                         & ChQM & $-1.25$ & $-0.16$ & $1.06$  &  $2.23$   & $-4.39$  \\
\noalign{\smallskip}
\multirow{2}{*}{2500}    & NQM  & $-0.71$ & $0.24$  & $1.74$  &  $2.14$   & $-4.84$  \\
\noalign{\smallskip}
                         & ChQM & $-1.27$ & $0.00$  & $1.10$  &  $1.97$   & $-4.34$  \\
\noalign{\smallskip}
\multirow{2}{*}{3000}    & NQM  & $-0.75$ & $0.18$  & $1.25$  &  $1.89$   & $-4.07$  \\
\noalign{\smallskip}
                         & ChQM & $-1.22$ & $0.10$  & $1.08$  &  $1.75$   & $-4.16$  \\
\noalign{\smallskip}
\multirow{2}{*}{3500}    & NQM  & $-0.81$ & $0.03$  & $0.55$  &  $1.70$   & $-3.10$ \\
\noalign{\smallskip}
                         & ChQM & $-1.16$ & $0.14$  & $0.93$  &  $1.57$   & $-3.80$ \\
\noalign{\smallskip}
\multirow{2}{*}{4000}    & NQM  & $-0.93$ & $-0.17$ & $-0.11$ &  $1.56$   & $-2.21$  \\
\noalign{\smallskip}
                         & ChQM & $-1.11$ & $0.13$  & $0.61$  &  $1.41$   & $-3.27$  \\
\noalign{\smallskip}
\multirow{2}{*}{4500}    & NQM  & $-1.04$ & $-0.20$ & $-0.22$ &  $1.38$   & $-2.00$  \\
\noalign{\smallskip}
                         & ChQM & $-1.10$ & $0.09$  & $0.11$  &  $1.29$   & $-2.59$  \\
\noalign{\smallskip}
\multirow{2}{*}{5000}    & NQM  & $-1.08$ & $-0.26$ & $-0.25$ &  $1.29$  & $-1.86$ \\
\noalign{\smallskip}
                         & ChQM & $-1.07$ & $0.04$  & $-0.07$ &  $1.18$  & $-2.22$ \\
\noalign{\smallskip}
\toprule[0.8pt]
\end{tabular}
\end{table}

{\it Dependence of binding mechanisms on the heavy quark mass}.
In order to clear the dependence of various mechanisms on the heavy quark mass,
we calculate the binding energy $E_b$ and various contributions to $E_b$ in the 
context of the heavy quark mass varying from 1500 MeV to 5000 MeV with a step
size of 500 MeV. The numerical results are presented in Table~\ref{dependence}.
One can find that the chromomagnetic term $\Delta V^{\rm cm}$ and kinetic energy
term $\Delta T$ dominant the properties of singled heavy flavor dibaryon states because
they directly depend on the heavy quark mass. Their signs do not change in the
range of heavy quark mass. Relatively speaking, the confinement term $\Delta V^{\rm con}$
and coulomb trem $\Delta V^{\rm coul}$ are weak and trivial for the formation of 
the bound dibaryon states. With the increase of heavy quark mass, the Coulomb term
$\Delta V^{\rm coul}$ is generally diminished while the confinement term $\Delta V^{\rm con}$
first increases and then decreases. The interval span range from $m_c$ to $m_b$ is
so large that the signs of each term are opposite in the di-$\Omega_{ccc}$ and
di-$\Omega_{bbb}$ states.

\section{summary}

In this work, we systematically investigate the single flavored dibaryons,
di-$\Delta^{++}$, di-$\Omega$, di-$\Omega_{ccc}$, and di-$\Omega_{bbb}$, with
$^1S_0$ in the quark models. In the calculation, we employ the Gaussian expansion
method, a high-precision numerical method. The di-$\Delta^{++}$, di-$\Omega_{ccc}$,
and di-$\Omega_{bbb}$ states can establish the deuteronlike bound state with a
binding energy about several MeV. However, the di-$\Omega$ state is a compact
deep bound state with a binding energy about 62 MeV.

Similar to chemical molecule covalent bond, the hadron covalent bond between two
colorless baryons can be established by the shared identical quarks induced by the
identical quark exchange effects. As a novel binding mechanism,
it plays a decisive role in the deuteronlike di-$\Omega_{ccc}$ and di-$\Omega_{bbb}$
states so that we call them covalent molecule states. Like the deuteron, the
$\sigma$-meson exchange play a dominant role in the light di-$\Delta^{++}$ and
di-$\Omega$ states. The hadron covalent bond clearly appears in the di-$\Delta^{++}$
state but is hidden in the di-$\Omega$ state by the larger relative motion kinetic
energy between two overlapped $\Omega$s. The chromomagnetic interaction is always
repulsive in the single flavored dibaryon states. The color-electric interaction
is strongly attractive in the di-$\Omega$ state but weakly attractive or repulsive
in the other dibayon states.

\acknowledgments{The author thanks Prof. S. L. Zhu for helpful discussions. This
research is supported by the Fundamental Research Funds for the Central Universities
under Contracts No. SWU118111.}

\end{document}